\documentclass{article}
\usepackage[accepted,nohyperref]{icml2020}
\usepackage[hidelinks,colorlinks=true,breaklinks]{hyperref}
\hypersetup{citecolor=[rgb]{0,0,0.5}}
\hypersetup{urlcolor=blue}
\hypersetup{linkcolor=[rgb]{0,0,0.5}}
\AtBeginShipout{%
  \ifnum\value{page}>1 %
    \typeout{* Additional boxing of page `\thepage'}%
    \setbox\AtBeginShipoutBox=\hbox{\copy\AtBeginShipoutBox}%
  \fi
}

\usepackage{microtype}
\usepackage{graphicx}
\usepackage{subfigure}
\usepackage{amsfonts}
\usepackage{amssymb}
\usepackage{amsthm}
\usepackage{amsmath}
\usepackage{comment}
\usepackage{booktabs} 

\usepackage[subtle]{savetrees}

\def \0bf{{\mathbf 0}}

\usepackage[newfloat,frozencache]{minted}
\usepackage[english]{babel}
\usepackage[T1]{fontenc}
\usepackage[utf8]{inputenc}
\usepackage{wallpaper}
\usepackage{palatino}
\usepackage[inline]{enumitem}
\setlist[enumerate,1]{label={(\arabic*)}}

\usepackage{graphicx}
\graphicspath{{figures/}}
\usepackage{amsmath,amssymb}

\usepackage{siunitx}
\usepackage{comment}

\usepackage[acronym,toc]{glossaries}
\setacronymstyle{long-short}

\usepackage{textcomp}

\DeclareSIUnit\carbon{\gram CO_{2}eq\per\kilo\watt\hour}
\DeclareSIUnit\gco{\gram CO_{2}eq}
\DeclareSIUnit\kgco{\kilo\gram CO_{2}eq}
\DeclareSIUnit\co{CO_{2}eq}
\DeclareSIUnit\gpud{GPU\text{-}days}
\newacronym[plural={GPUs},\glsshortpluralkey={GPUs}]{gpu}{GPU}{graphics processing unit}
\newacronym[plural={CPUs},\glsshortpluralkey={CPUs}]{cpu}{CPU}{central processing unit}
\newacronym[plural={TPUs},\glsshortpluralkey={TPUs}]{tpu}{TPU}{tensor processing unit}
\newacronym{dram}{DRAM}{dynamic random-access memory}
\newacronym{dl}{DL}{deep learning}
\newacronym{ml}{ML}{machine learning}
\newacronym{sota}{SOTA}{state-of-the-art}
\newacronym[plural={DNNs},\glsshortpluralkey={DNNs}]{dnn}{DNN}{deep neural network}
\newacronym[plural={CNNs},\glsshortpluralkey={CNNs}]{cnn}{CNN}{deep convolutional neural network}
\newacronym{nlp}{NLP}{natural language processing}
\newacronym[plural={NNs},\glsshortpluralkey={NNs}]{nn}{NN}{neural network}
\newacronym{pue}{PUE}{Power Usage Effectiveness}
\newacronym{ghg}{GHG}{greenhouse gas}
\newacronym[plural={RECs},\glsshortpluralkey={RECs}]{rec}{REC}{Renewable Energy Credit}
\newacronym{pypi}{PyPi}{the Python Package Index}
\newacronym{tdp}{TDP}{thermal design power}
\newacronym{ai}{AI}{artificial intelligence}
\newacronym{fpos}{FPOs}{floating point operations}
\newacronym[plural={APIs},\glsshortpluralkey={APIs}]{api}{API}{application programming interface}
\newacronym{nvml}{NVML}{NVIDIA Management Library}
\newacronym{rapl}{Intel RAPL}{Intel Running Average Power Limit}
\newacronym{flops}{FLOPS}{floating point operations per second}
\newacronym{dvfs}{DVFS}{dynamic voltage and frequency scaling}

\begin{document}
\glsdisablehyper

\twocolumn[
\icmltitle{Carbontracker: Tracking and Predicting the Carbon Footprint of Training Deep Learning Models}
\icmltitlerunning{Carbontracker}




\begin{icmlauthorlist}
\icmlauthor{Lasse F. Wolff Anthony$^*$}{ku}
\icmlauthor{Benjamin Kanding$^*$}{ku}
\icmlauthor{Raghavendra Selvan}{ku}

\end{icmlauthorlist}
\icmlaffiliation{ku}{Department of Computer Science, University of Copenhagen, Copenhagen, Denmark}
\icmlcorrespondingauthor{Lasse F. Wolff Anthony}{lassewolffanthony@gmail.com}
\icmlcorrespondingauthor{Benjamin Kanding}{bmk1212@live.dk}
\icmlkeywords{Machine Learning, ICML, Deep Learning, Carbon Footprint, Carbon Emissions, Climate Change, Energy Consumption, Energy-efficiency, GPUs}

\vskip 0.3in
]


\printAffiliationsAndNotice{\icmlEqualContribution} 

\begin{abstract}
Deep learning (DL) can achieve impressive results across a wide variety of tasks, but this often comes at the cost of training models for extensive periods on specialized hardware accelerators. This energy-intensive workload has seen immense growth in recent years. Machine learning (ML) may become a significant contributor to climate change if this exponential trend continues. If practitioners are aware of their energy and carbon footprint, then they may actively take steps to reduce it whenever possible. In this work, we present {\em carbontracker}, a tool for tracking and predicting the energy and carbon footprint of training DL models. We propose that energy and carbon footprint of model development and training is reported alongside performance metrics using tools like {\em carbontracker}. We hope this will promote responsible computing in ML and encourage research into energy-efficient deep neural networks.~\footnote{Source code for {\em carbontracker} is available here: \url{https://github.com/lfwa/carbontracker}} 
\end{abstract}

\section{Introduction}

The popularity of solving problems using \gls{dl} has rapidly increased and with it the need for ever more powerful models. These models achieve impressive results across a wide variety of tasks such as gameplay, where AlphaStar reached the highest rank in the strategy game Starcraft II \cite{vinyals2019grandmaster} and Agent57 surpassed human performance in all 57 Atari 2600 games \cite{badia2020agent57}. This comes at the cost of training the model for thousands of hours on specialized hardware accelerators such as \glspl{gpu}. From 2012 to 2018 the compute needed for \gls{dl} grew \SI{300000}{}-fold \cite{amodei2018ai}.

This immense growth in required compute has a high energy demand, which in turn increases the demand for energy production. In 2010 energy production was responsible for approximately $35$\% of total anthropogenic \gls{ghg} emissions \cite{ipcc2014}. Should this exponential trend in \gls{dl} compute continue then \gls{ml} may become a significant contributor to climate change.

This can be mitigated by exploring how to improve energy efficiency in \gls{dl}. Moreover, if practitioners are aware of their energy and carbon footprint, then they may actively take steps to reduce it whenever possible. We show that in \gls{ml}, these can be simple steps that result in considerable reductions to carbon emissions. 

The environmental impact of \gls{ml} in research and industry has seen increasing interest in the last year following the 2018 IPCC special report \cite{Masson-Delmotte2018} calling for urgent action in order to limit global warming to \SI{1.5}{\celsius}. We briefly review some notable work on the topic.
\citet{Strubell2019} estimated the financial and environmental costs of R\&D and hyperparameter tuning for various \gls{sota} \gls{nn} models in \gls{nlp}. They point out that increasing cost and emissions of \gls{sota} models contribute to a lack of equity between those researchers who have access to large-scale compute, and those who do not. The authors recommend that metrics such as training time, computational resources required, and model sensitivity to hyperparameters should be reported to enable direct comparison between models. 
\citet{Lacoste} provided the \textit{Machine Learning Emissions Calculator}
that relies on self-reporting. The tool can estimate the carbon footprint of \gls{gpu} compute by specifying hardware type, hours used, cloud provider, and region. \citet{Henderson2020} presented the \textit{experiment-impact-tracker} 
framework and gave various strategies for mitigating carbon emissions in \gls{ml}. Their Python framework allows for estimating the energy and carbon impact of \gls{ml} systems as well as the generation of ``Carbon Impact Statements'' for standardized reporting hereof.

In this work, we propose \textit{carbontracker}, a tool for tracking and predicting the energy consumption and carbon emissions of training \gls{dl} models. The methodology is similar to that of \citet{Henderson2020} but differs from prior art in two major ways:
\begin{enumerate}
    \item We allow for a further proactive and intervention-driven approach to reducing carbon emissions by supporting predictions. Model training can be stopped, at the user's discretion, if the predicted environmental cost is exceeded.
    \item We support a variety of different environments and platforms such as clusters, desktop computers, and Google Colab notebooks, allowing for a plug-and-play experience.
\end{enumerate}

We experimentally evaluate the tool on several different \gls{cnn} architectures and datasets for medical image segmentation and assess the accuracy of its predictions. We present concrete recommendations on how to reduce carbon emissions considerably when training \gls{dl} models.

\section{Design and Implementation}\label{sec:design_and_implementation}

The design philosophy that guided the development of \textit{carbontracker} can be summarized by the following principles:

\begin{description}[wide=0\parindent,itemsep=0pt]
    \item[Pythonic]
    The majority of \gls{ml} takes place in the Python language \cite{visionmobile2017}. We want the tool to be as easy as possible to integrate into existing work environments making Python the language of choice. 
    
    \item[Usable]
    The required effort and added code must be minimal and not obfuscate the existing code structure.
    
    \item[Extensible]
    Adding and maintaining support for changing \glspl{api} and new hardware should be straightforward.
    
    \item[Flexible]
    The user should have full control over what is monitored and how this monitoring is performed.
    
    \item[Performance]
    The performance impact of using the tool must be negligible, and computation should be minimal. It must not affect training.
    
    \item[Interpretable]
    Carbon footprint expressed in \si{\gco} is often meaningless. A common understanding of the impact should be facilitated through conversions.
    
\end{description}

\textit{Carbontracker} is an open-source tool written in Python for tracking and predicting the energy consumption and carbon emissions of training \gls{dl} models. It is available through \gls{pypi}.
The tool is implemented as a multithreaded program.
It utilizes separate threads to collect power measurements and fetch carbon intensity in real-time for parallel efficiency and to not disrupt the model training in the main thread. \autoref{app:implementation} has further implementation details.

\textit{Carbontracker} supports predicting the total duration, energy, and carbon footprint of training a \gls{dl} model. These predictions are based on a user-specified number of monitored epochs with a default of 1.
We forecast the carbon intensity of electricity production during the predicted duration using the supported \glspl{api}. The forecasted carbon intensity is then used to predict the carbon footprint. Following our preliminary research, we use a simple linear model for predictions.

\section{Experiments and Results}
\label{chap:experiments}
In order to evaluate the performance and behavior of \textit{carbontracker}, we conducted experiments on three medical image datasets using two different \gls{cnn} models: U-net \cite{Ronneberger2015} and lungVAE \cite{Selvan2020}.  The models were trained for the task of medical image segmentation using three datasets: DRIVE \cite{Staal2004}, LIDC \cite{armato2004lung}, and CXR \cite{jaeger2014two}. Details on the models and datasets are given in \autoref{sec:models_and_data}. All measurements were taken using \textit{carbontracker} version 1.1.2. We performed our experiments on a single NVIDIA TITAN RTX \gls{gpu} with $12$ GB memory and two Intel \glspl{cpu}.

In line with our message of reporting energy and carbon footprint, we used \textit{carbontracker} to generate the following statement: 
The training of models in this work is estimated to use \SI{37.445}{\kilo\watt\hour} of electricity contributing to \SI{3.166}{\kilo\gram} of \si{\co}. This is equivalent to \SI{26.296}{\kilo\meter} travelled by car (see \ref{logging}).

\begin{figure}[t]
    \centering
    \includegraphics[width=1\linewidth]{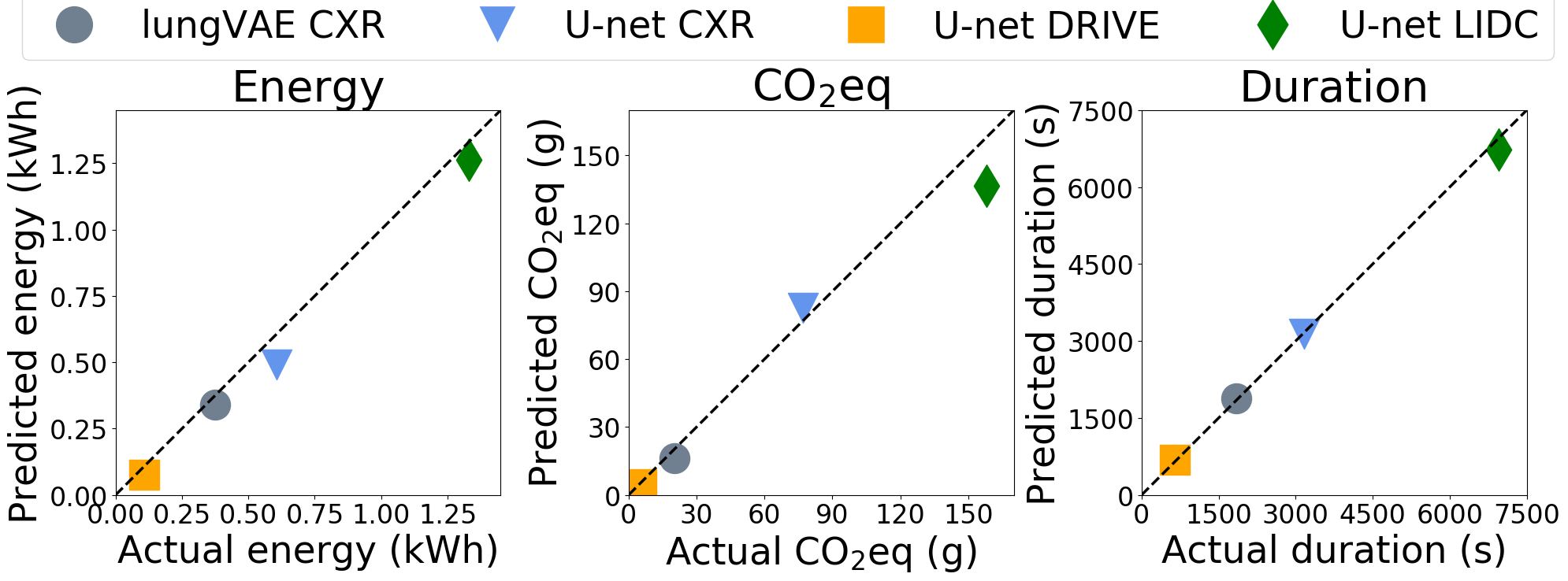}
    \caption{Comparison of predicted and measured values of energy in \SI{}{\kilo\watt\hour} (left), emissions in \SI{}{\gco} (center), and duration in \SI{}{\second} (right) for the full training session when predicting after a single epoch. The diagonal line represents predictions that are equal to the actual measured consumption. 
    Description of the models and datasets are in  \autoref{sec:models_and_data}.}
    \label{fig:pred_vs_actual}
\end{figure}

\begin{figure}[t]
    \centering
    \includegraphics[width=0.8\linewidth]{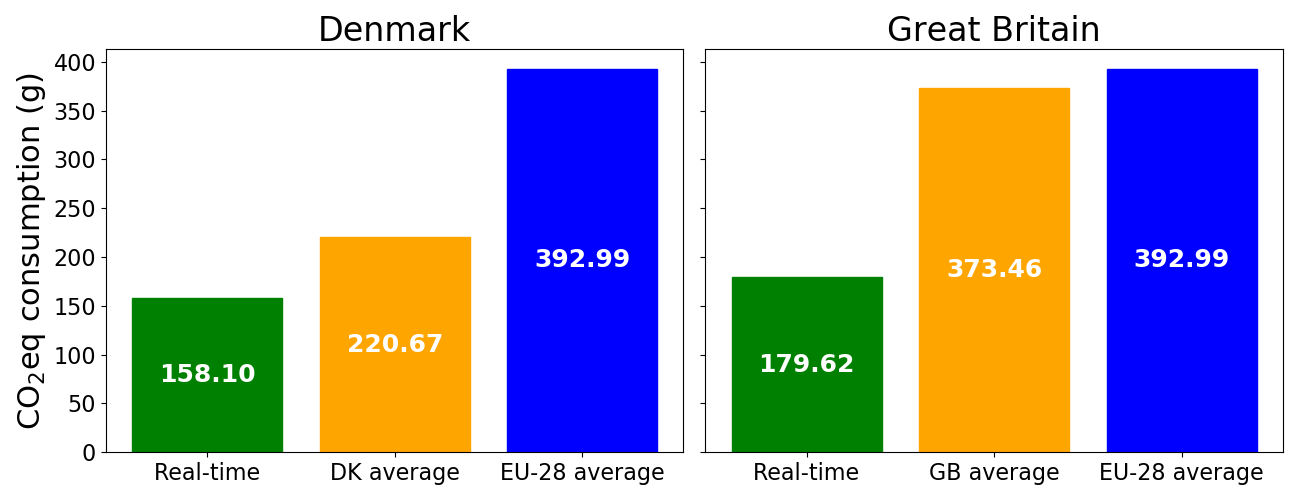}
    \caption{Carbon emissions (\SI{}{\gco}) of training the U-net on LIDC dataset for different carbon intensity estimation methods. (left) The emissions of training in Denmark and (right) in Great Britain at 2020-05-21 22:00 local time. Real-time indicates that the current intensity is fetched every \SI{15}{\minute} during training using the \glspl{api} supported by \textit{carbontracker}. The average intensities are from 2016 (see \autoref{fig:carbonintensity_europe} in Appendix).}
    \label{fig:realtime_importance}
    \vspace{-1em}
\end{figure}

An overview of predictions from \textit{carbontracker} based on monitoring for $1$ training epoch for the trained models compared to the measured values is shown in \autoref{fig:pred_vs_actual}. The errors in the energy predictions are $4.9$--$19.1$\% compared to the measured energy values, $7.3$--$19.9$\% for the \si{\co}, and $0.8$--$4.6$\% for the duration. The error in the \si{\co} predictions are also affected by the quality of the forecasted carbon intensity from the \glspl{api} used by \textit{carbontracker}. This is highlighted in \autoref{fig:realtime_importance}, which shows the estimated carbon emissions (\si{\gco}) of training our U-net model on LIDC in Denmark and Great Britain for different carbon intensity estimation methods. As also shown by \citet{Henderson2020}, we see that using country or region-wide average estimates may severely overestimate (or under different circumstances underestimate) emissions. This illustrates the importance of using real-time (or forecasted) carbon intensity for accurate estimates of carbon footprint.


\begin{figure}[t]
    \centering
    \includegraphics[width=0.7\linewidth]{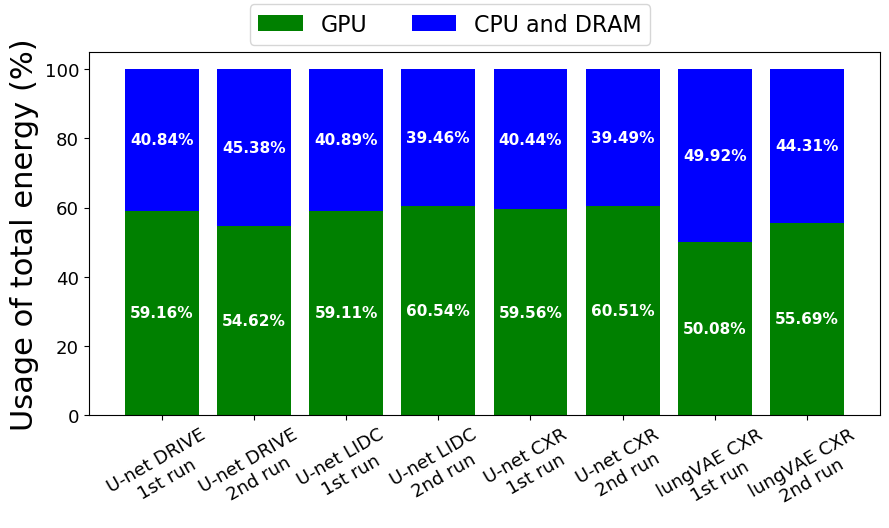}
    \caption{Comparison of energy usage by component shown as the relative energy usage (\%) out of the total energy spent during training. We see that the \gls{gpu} uses the majority of the energy, about $50$--$60$\%, but the \gls{cpu} and \gls{dram} also account for a significant amount of the total energy consumption across all models and datasets.}
    \label{fig:relative_component_usage}
\end{figure}

\autoref{fig:relative_component_usage} summarizes the relative energy consumption of each component across all runs. We see that while the \gls{gpu} uses the majority of the total energy, around $50$--$60$\%, the \gls{cpu} and \gls{dram} also account for a significant part of the total consumption. This is consistent with the findings of \citet{Gorkovenko2020}, who found that \glspl{gpu} are responsible for around $70$\% of power consumption, \gls{cpu} for $15$\%, and RAM for $10$\% when testing on the TensorFlow benchmarking suite for \gls{dl} on Lenovo ThinkSystem SR670 servers. As such, only accounting for \gls{gpu} consumption when quantifying the energy and carbon footprint of \gls{dl} models will lead to considerable underestimation of the actual footprint.


\section{Reducing Your Carbon Footprint}\label{chap:reduce_footprint}

The carbon emissions that occur when training \gls{dl} models are not irreducible and do not have to simply be the cost of progress within \gls{dl}. Several steps can be taken in order to reduce this footprint considerably. In this section, we outline some strategies for practitioners to directly mitigate their carbon footprint when training \gls{dl} models.

\begin{figure}[t]
    \centering
    \includegraphics[width=0.935\linewidth]{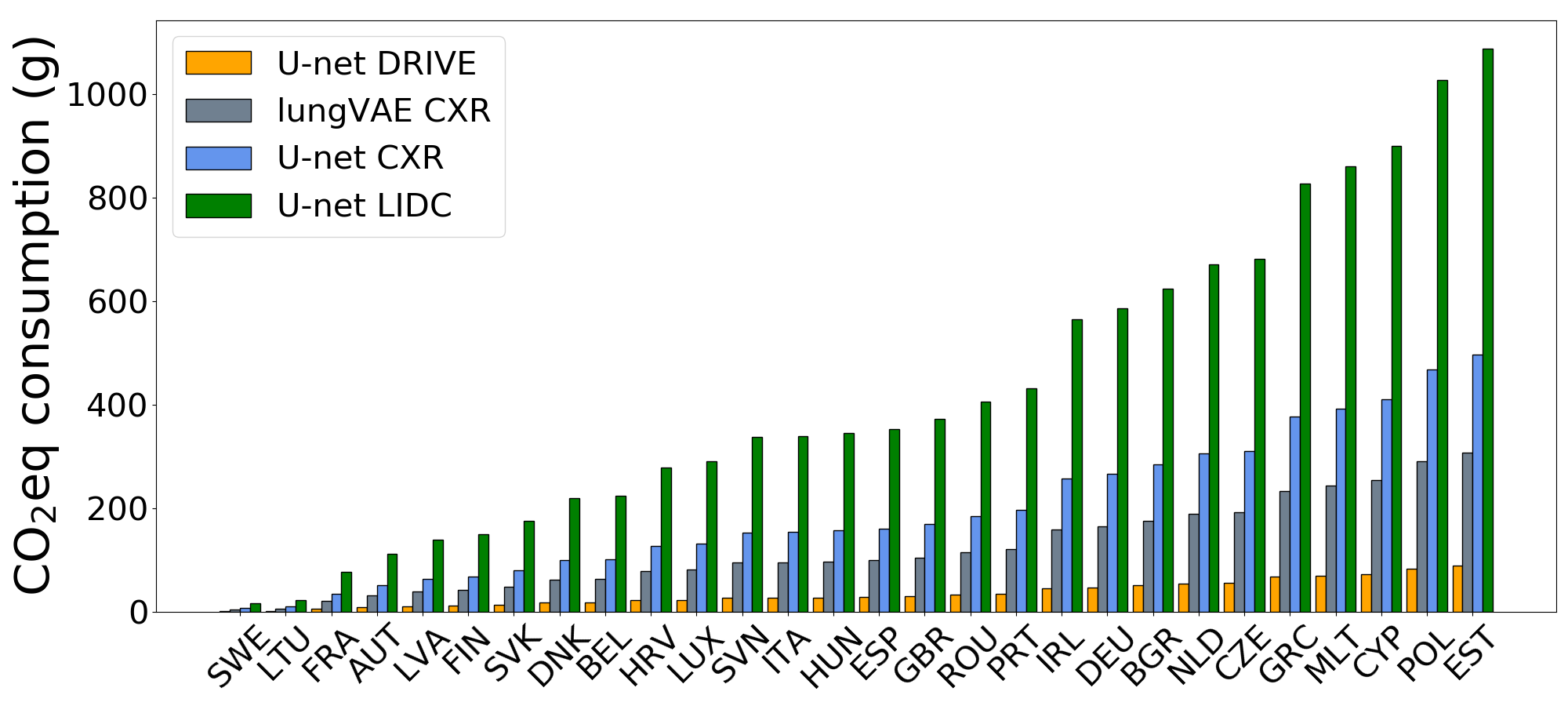}
    \caption{Estimated carbon emissions (\SI{}{\gco}) of training our models (see \autoref{sec:models_and_data}) in different EU-28 countries. The calculations are based on the average carbon intensities from 2016 (see \autoref{fig:carbonintensity_europe} in Appendix).}
    \label{fig:region_savings}
    \vspace{-1em}
\end{figure}

\begin{description}[wide=0\parindent,itemsep=0pt]

\item[Low Carbon Intensity Regions]

The carbon intensity of electricity production varies by region and is dependent on the energy sources that power the local electrical grid. 
\autoref{fig:region_savings} illustrates how the variation in carbon intensity between regions can influence the carbon footprint of training \gls{dl} models. Based on the 2016 average intensities, we see that a model trained in Estonia may emit more than 61 times the \si{\co} as an equivalent model would when trained in Sweden. In perspective, our U-net model trained on the LIDC dataset would emit \SI{17.7}{\gco} or equivalently the same as traveling \SI{0.14}{\kilo\meter} by car when trained in Sweden. However, training in Estonia it would emit \SI{1087.9}{\gco} or the same as traveling \SI{9.04}{\kilo\meter} by car for just a single training session.

As training \gls{dl} models is generally not latency bound, we recommend that \gls{ml} practitioners move training to regions with a low carbon intensity whenever it is possible to do so. We must further emphasize that for large-scale models that are trained on multiple \glspl{gpu} for long periods, such as OpenAI's GPT-3 language model \cite{brown2020language}, it is imperative that training takes place in low carbon intensity regions in order to avoid several megagrams of carbon emissions. The absolute difference in emissions may even be significant between two green regions, like Sweden and France, for such large-scale runs.


\item[Training Times]

\begin{figure}[t]
    \centering
    \includegraphics[width=0.85\linewidth]{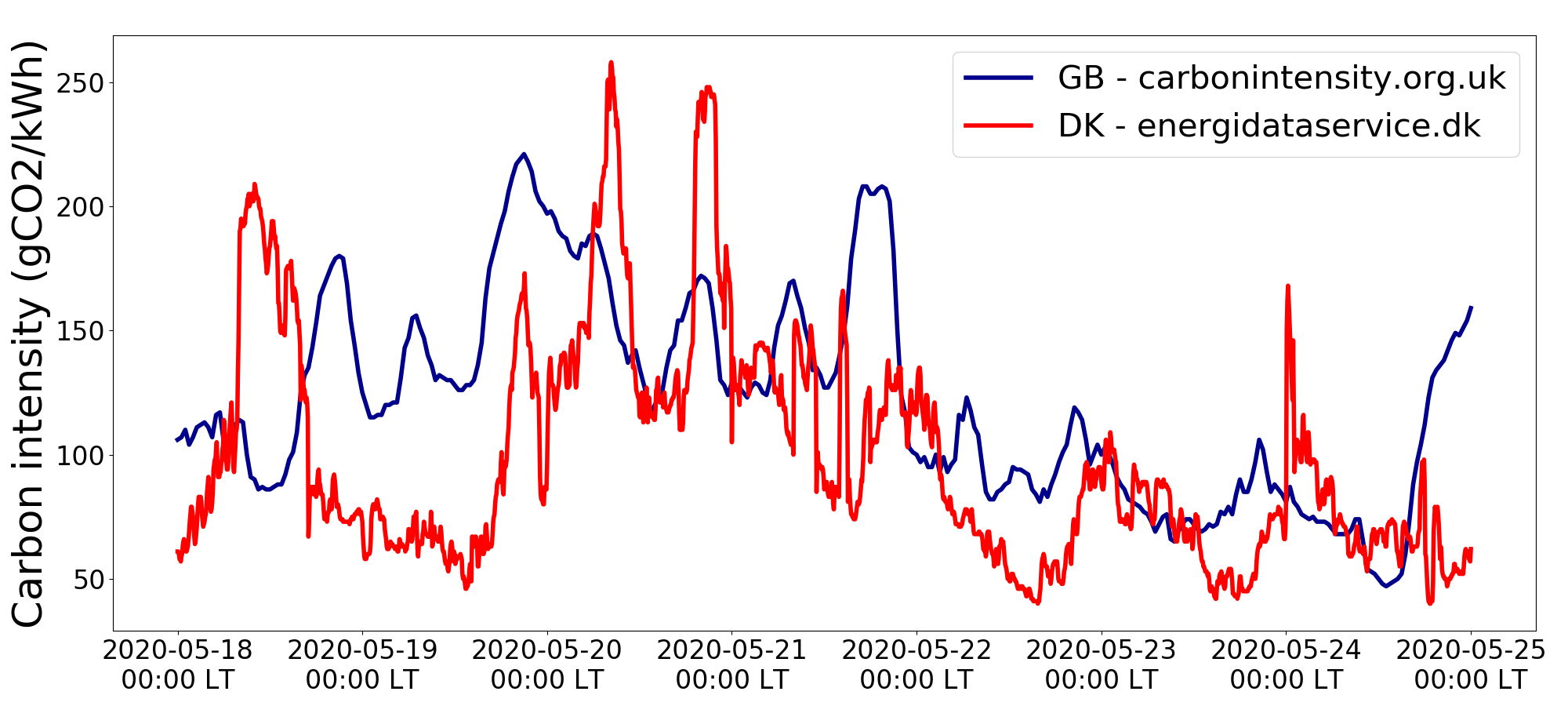}
    \vspace{-0.25em}
    \caption{Real-time carbon intensity (\SI{}{\carbon}) for Denmark (DK) and Great Britain (GB) from 2020-05-18 to 2020-05-25 shown in local time. The data is collected using the \glspl{api} supported by \textit{carbontracker}. The carbon intensities are volatile to changes in energy demand and depend on the energy sources available.}
    \label{fig:carbon_intensities}
    \vspace{-0.75em}
\end{figure}

The time period in which a \gls{dl} model is trained affects its overall carbon footprint. This is caused by carbon intensity changing throughout the day as energy demand and capacity of energy sources change. \autoref{fig:carbon_intensities} shows the carbon intensity (\si{\carbon}) for Denmark and Great Britain in the week of 2020-05-18 to 2020-05-25 collected with the \glspl{api} supported by \textit{carbontracker}. A model trained during low carbon intensity hours of the day in Denmark may emit as little as $\frac{1}{4}$ the \si{\co} of one trained during peak hours. A similar trend can be seen for Great Britain, where $2$-fold savings in emissions can be had. 

We suggest that \gls{ml} practitioners shift training to take place in low carbon intensity time periods whenever possible. The time period should be determined on a regional level.

\item[Efficient Algorithms]
The use of efficient algorithms when training \gls{dl} models can further help reduce compute-resources and thereby also carbon emissions. Hyperparameter tuning may be improved by substituting grid search for random search \cite{Bergstra2012}, using Bayesian optimization \cite{snoek2012practical} or other optimization techniques like Hyperband \cite{hyperband2017}. Energy efficiency of inference in \glspl{dnn} is also an active area of research with methods such as quantization aware training, energy-aware pruning \cite{Yang2017}, and power- and memory-constrained hyperparameter optimization like HyperPower \cite{Stamoulis2018}. 

\item[Efficient Hardware and Settings]
Choosing more energy-efficient computing hardware and settings may also contribute to reducing carbon emissions. Some \glspl{gpu} have substantially higher efficiency in terms of \gls{flops} per watt of power usage compared to others \cite{Lacoste}.
Power management techniques like \gls{dvfs} can further help conserve energy consumption \cite{Li2016} and for some models even reduce time to reach convergence \cite{Tang2019}. \citet{Tang2019} show that \gls{dvfs} can be applied to \glspl{gpu} to help conserve about $8.7$\% to $23.1$\% energy consumption for training different \glspl{dnn} and about $19.6$\% to $26.4$\% for inference. Moreover, the authors show that the default frequency settings on tested \glspl{gpu}, such as NVIDIA's Pascal P100 and Volta V100, are often not optimized for energy efficiency in \gls{dnn} training and inference.

\end{description}

\vspace{-0.5cm}
\section{Discussion and Conclusion}

The current trend in \gls{dl} is a rapidly increasing demand for compute that does not appear to slow down. This is evident in recent models such as the GPT-3 language model \cite{brown2020language} with $175$ billion parameters requiring an estimated \SI{28000}{\gpud} to train excluding R\&D (see \autoref{sec:appendix_gpt3}). We hope to spread awareness about the environmental impact of this increasing compute through accurate reporting with the use of tools such as \textit{carbontracker}. Once informed, concrete and often simple steps can be taken in order to reduce the impact.

\Gls{sota}-results in \gls{dl} are frequently determined by a model's performance through metrics such as accuracy, AUC score, or similar performance metrics. Energy-efficiency is usually not one of these. While such performance metrics remain a crucial measure of model success, we hope to promote an increasing focus on energy-efficiency.
We must emphasize that we do not argue that compute-intensive research is not essential for the progress of \gls{dl}. We believe, however, that the impact of this compute should be minimized.
We propose that the total energy and carbon footprint of model development and training is reported alongside accuracy and similar metrics to promote responsible computing in \gls{ml} and research into energy-efficient \glspl{dnn}.

In this work, we showed that \gls{ml} risks becoming a significant contributor to climate change. To this end, we introduced the open-source \textit{carbontracker} tool for tracking and predicting the total energy consumption and carbon emissions of training \gls{dl} models. This enables practitioners to be aware of their footprint and take action to reduce it.

\subsubsection*{Acknowledgements}
The authors would like to thank Morten Pol Engell-Nørregård for the thorough feedback on the thesis version of this work. The authors also thank the anonymous reviewers and early users of \textit{carbontracker} for their insightful feedback.

\bibliography{references}
\bibliographystyle{icml2020}
\appendix

\section{Implementation details}
\label{app:implementation}

\begin{listing}[h]
\caption{Example of the default setup added to training scripts for tracking and predicting with \textit{carbontracker}.}
\begin{minted}[linenos,tabsize=2,breaklines,fontsize=\footnotesize]{python}
from carbontracker.tracker import CarbonTracker


tracker = CarbonTracker(epochs=<your epochs>)

for epoch in range(<your epochs>):
    tracker.epoch_start()

    # Your model training.

    tracker.epoch_end()

tracker.stop()
\end{minted}
\label{code:minimal_setup}
\end{listing}
\begin{listing}[t]
\caption{Example output of using \textit{carbontracker} to track and predict the energy and carbon footprint of training a \gls{dl} model.}
\begin{minted}[breaklines,fontsize=\footnotesize]{console}
CarbonTracker: The following components were found: GPU with device(s) TITAN RTX. CPU with device(s) cpu:0, cpu:1.
CarbonTracker: Carbon intensity for the next 1:54:54 is predicted to be 54.09 gCO2/kWh at detected location: Copenhagen, Capital Region, DK.
CarbonTracker: 
Predicted consumption for 100 epoch(s):
	Time:   1:54:54
	Energy: 1.159974 kWh
	CO2eq:  62.744032 g
	This is equivalent to:
	0.521130 km travelled by car
CarbonTracker: Average carbon intensity during training was 58.25 gCO2/kWh at detected location: Copenhagen, Capital Region, DK.
CarbonTracker: 
Actual consumption for 100 epoch(s):
	Time:   1:55:55
	Energy: 1.334319 kWh
	CO2eq:  77.724065 g
	This is equivalent to:
	0.645549 km travelled by car
CarbonTracker: Finished monitoring.
\end{minted}
\label{code:example_output}
\end{listing}

\begin{figure*}[t]
    \centering
    \includegraphics[width=0.65\linewidth]{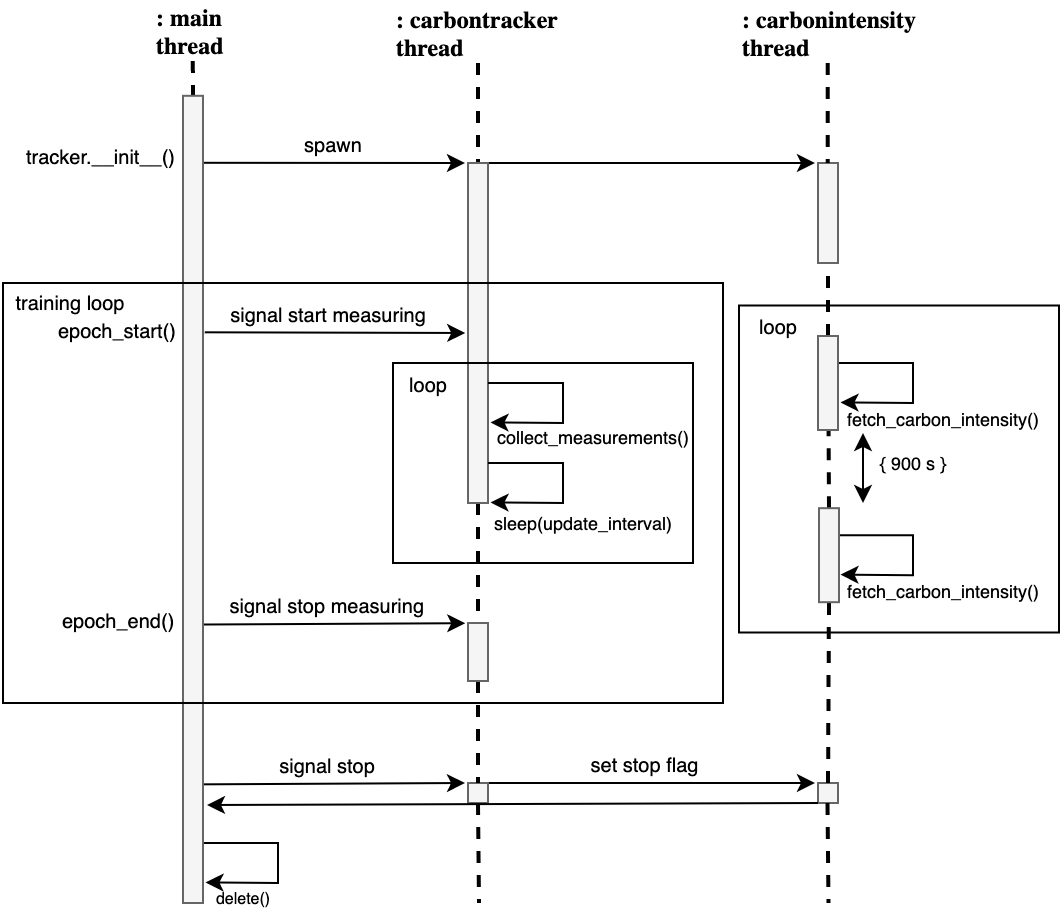}
    \caption{A visualization of the \textit{carbontracker} control flow. The \texttt{main} thread instantiates the \texttt{CarbonTracker} class which then spawns the \texttt{carbontracker} and \texttt{carbonintensity} daemon threads. The \texttt{carbontracker} thread continuously collects power measurements for available devices. The \texttt{carbonintensity} thread fetches the current carbon intensity every \SI{900}{\second}. When the specified epochs before predicting have passed, the total predicted consumption is reported to \texttt{stdout} and optional log files. Similarly, when the specified amount of epochs have been monitored, the actual measured consumption is reported, after which the \texttt{carbontracker} and \texttt{carbonintensity} threads join the \texttt{main} thread. Finally, the \textit{carbontracker} object runs the cleanup routine \texttt{delete()} which releases all used resources.}
    \label{fig:carbontracker}
\end
{figure*}

\textit{Carbontracker} is a multithreaded program. \autoref{fig:carbontracker} illustrates a high-level overview of the program.

\subsection{On the Topic of Power and Energy Measurements}\label{sec:on_the_topic_of_power_and_energy}
In our work, we measure the total power of selected components such as the \gls{gpu}, \gls{cpu}, and \gls{dram}. It can be argued that dynamic power rather than total power would more fairly represent a user's power consumption when using large clusters or cloud computing. We argue that these computing resources would not have to exist if the user did not use them.
As such, the user should also be accountable for the static power consumption during the period in which they reserve the resource. It is also a pragmatic solution as accurately estimating dynamic power is challenging due to the infeasibility of measuring static power by software and the difficulty in storing and updating information about static power for a multitude of different components. 
A similar argument can be made for the inclusion of life-cycle aspects in our energy estimates, such as accounting for the energy attributed to the manufacturing of system components. Like \citet{Henderson2020}, we ignore these aspects due to the difficulties in their estimation.

The power and energy monitoring in \textit{carbontracker} is limited to a few main components of computational systems. Additional power consumed by the supporting infrastructure, such as that used for cooling or power delivery, is accounted for by multiplying the measured power by the \gls{pue} of the data center hosting the compute, as suggested by \citet{Strubell2019}. \Gls{pue} is a ratio describing the efficiency of a data center and the energy overhead of the computing equipment. It is defined as the ratio of the total energy used in a data center facility to the energy used by the IT equipment such as the compute, storage, and network equipment \cite{Avelar2012}:
\begin{equation}
    \text{PUE} = \frac{\text{Total Facility Energy}}{\text{IT Equipment Energy}}.
\end{equation}
Previous research has examined \gls{pue} and its shortcomings~\cite{Yuventi2013}. These shortcomings may largely be resolved by data centers reporting an average \gls{pue} instead of a minimum observed value. In our work, we use a \gls{pue} of $1.58$, the global average for data centers in 2018 as reported by \citet{Ascierto2018}.\footnote{Early versions (v1.1.2 and earlier) of \textit{carbontracker} used a \gls{pue} of $1.58$. This has subsequently been updated to $1.67$, the global average for 2019 \cite{Ascierto2019}.} This may lead to inaccurate estimates of power and energy consumption of compute in energy-efficient data centers; e.g., Google reports a fleetwide \gls{pue} of $1.10$ for 2020.\footnote{See \url{https://www.google.com/about/datacenters/efficiency/}} Future work may, therefore, explore alternatives to using an average \gls{pue} value as well as to include more components in our measurements to improve the accuracy of our estimation. Currently, the software needed to take power measurements for components beyond the \gls{gpu}, \gls{cpu}, and \gls{dram} is extremely limited or in most cases, non-existent.

\subsection{On the Topic of Carbon Offsetting}
Carbon emissions can be compensated for by carbon offsetting or the purchases of \glspl{rec}. Carbon offsetting is the reduction in emissions made to compensate for emissions occurring elsewhere \cite{goodward2010bottom}. We ignore such offsets and \glspl{rec} in our reporting as to encourage responsible computing in the \gls{ml} community that would further reduce global emissions. See \citet{Henderson2020} for an extended discussion on carbon offsets and why they also do not account for them in the \textit{experiment-impact-tracker} framework. Our carbon footprint estimate is solely based on the energy consumed during training of \gls{dl} models.

\subsection{Power and Energy Tracking}\label{sec:power_and_energy_monitoring}
The power and energy tracking in \textit{carbontracker} occurs in the \texttt{carbontracker} thread. The thread continuously collects instantaneous power samples in real-time for every available device of the specified components. Once samples for every device has been collected, the thread will sleep for a fixed interval before collecting samples again. When the epoch ends, the thread stores the epoch duration. Finally, after the training loop completes we calculate the total energy consumption $E$ as
\begin{equation}\label{eq:compute_energy}
    E = \text{PUE} \sum_{e \in \mathcal{E}} \sum_{d \in \mathcal{D}} P_{avg,de} T_e
\end{equation}
where $P_{avg,de}$ is the average power consumed by device $d \in \mathcal{D}$ in epoch $e \in \mathcal{E}$, and $T_e$ is the duration of epoch $e$.


The components supported by \textit{carbontracker} in its current form are the \gls{gpu}, \gls{cpu}, and \gls{dram} due to the aforementioned restrictions. NVIDIA \glspl{gpu} represent a large share of Infrastructure-as-a-Service compute instance types with dedicated accelerators.
So we support NVIDIA \glspl{gpu} as power sampling is exposed through the \gls{nvml}\footnote{\url{https://developer.nvidia.com/nvidia-management-library-nvml}}. 
Likewise, we support Intel \glspl{cpu} and \gls{dram} through the \gls{rapl} interface \cite{David2010}.

\subsection{Converting Energy Consumption to Carbon Emissions}

We can estimate the carbon emissions resulting from the electricity production of the energy consumed during training as the product of the energy and carbon intensity as shown in \eqref{eq:carbon_footprint}:
\begin{align}
    \text{Carbon Footprint} = & \text{ Energy Consumption} \times \nonumber \\ &\text{ Carbon Intensity}.
    \label{eq:carbon_footprint}
\end{align}
The used carbon intensity heavily influences the accuracy of this estimate. In \textit{carbontracker}, we support the fetching of carbon intensity in real-time through external \glspl{api}.We dynamically determine the location based on the IP address of the local compute through the Python geocoding library \texttt{geocoder}\footnote{\url{https://github.com/DenisCarriere/geocoder}}. 
Unfortunately, there does not currently exist a globally accurate, free, and publicly available real-time carbon intensity database. This makes determining the carbon intensity to use for the conversion more difficult. 

We solve this problem by using several \glspl{api} that are local to each region. It is currently limited to Denmark and Great Britain. Other regions default to an average carbon intensity for the EU-28 countries in 2017\footnote{\url{https://www.eea.europa.eu/data-and-maps/data/co2-intensity-of-electricity-generation}}. For Denmark we use data from Energi Data Service\footnote{\url{https://energidataservice.dk/}} and for Great Britain we use the Carbon Intensity API\footnote{\url{https://carbonintensity.org.uk/}}.

\subsection{Logging}\label{logging}
Finally, \textit{carbontracker} has extensive logging capabilities enabling transparency of measurements and enhancing the reproducibility of experiments. The user may specify the desired path for these log files. We use the logging API\footnote{\url{https://docs.python.org/3/library/logging.html}} provided by the standard Python library. 

Additional functionality for interaction with logs has also been added through the \texttt{carbontracker.parser} module. Logs may easily be parsed into Python dictionaries containing all information regarding the training sessions, including power and energy usages, epoch durations, devices monitored, whether the model stopped early, and the outputted prediction.
We further support aggregating logs into a single estimate of the total impact of all training sessions.
By using different log directories, the user can easily keep track of the total impact of each model trained and developed. The user may then use the provided parser functionality to estimate the full impact of R\&D.

Current version of \textit{carbontracker} uses kilometers travelled by car as the carbon emissions conversion. This data is retrieved from the average \si{\co} emissions of a newly registered car in the European Union in 2018\footnote{\url{https://www.eea.europa.eu/data-and-maps/indicators/average-co2-emissions-from-motor-vehicles/assessment-1}}.

\section{Models and Data}\label{sec:models_and_data}
In our experimental evaluation, we trained two \gls{cnn} models on three medical image datasets for the task of image segmentation. The models were developed in PyTorch \cite{Paszke2019}. We describe each of the models and datasets in turn below.
\begin{description}[wide=0\parindent]
    \item[U-net DRIVE]
    This is a standard U-net model \cite{Ronneberger2015} trained on the DRIVE dataset \cite{Staal2004}. DRIVE stands for Digital Retinal Images for Vessel Extraction and is intended for segmentation of blood vessels in retinal images. The images are $768$ by $584$ pixels and JPEG compressed. We used a training set of $15$ images and trained for $300$ epochs with a batch size of $4$ and a learning rate of $10^{-3}$ with the Adam optimizer \cite{kingma2014adam}.

    \item[U-net CXR]
    The model is based on a U-net \cite{Ronneberger2015} with slightly changed parameters. The dataset comprises of chest X-rays (CXR) with lung masks curated for pulmonary tuberculosis detection \cite{jaeger2014two}. We use $528$ CXRs for training and $176$ for validation without any data augmentation. We trained the model for $200$ epochs with a batch size of $12$, a learning rate of $10^{-4}$, and weight decay of $10
   ^{-5}$ with the Adam optimizer \cite{kingma2014adam}.
    
    \item[U-net LIDC]
    This is also a standard U-net model \cite{Ronneberger2015} but trained on a preprocessed LIDC-IDRI dataset~\cite{armato2004lung}\footnote{\url{https://github.com/stefanknegt/Probabilistic-Unet-Pytorch}}. The LIDC-IDRI dataset consists of 1018 thoracic computed tomography (CT) scans with annotated lesions from four different radiologists. We trained our model on the annotations of a single radiologist for $100$ epochs. We used a batch size of $64$ and a learning rate of $10^{-3}$ with the Adam optimizer \cite{kingma2014adam}.

    \item[lungVAE CXR]
    This model and dataset is from the open source model available from \citet{Selvan2020}. The model uses a U-net type segmentation network and a variational encoder for data imputation. The dataset is the same CXR dataset as used in our above U-net CXR model. $528$ CXRs are used for training and $176$ for validation. The model was trained with a batch size of $12$ and a learning rate of $10^{-4}$ with the Adam optimizer \cite{kingma2014adam} for a maximum of $200$ epoch using early stopping based on the validation loss. The first run was $90$ epochs, and the second run was $97$.
\end{description}

\section{Additional Experiments}

\begin{figure}[h!]
    \centering
    \includegraphics[width=0.75\linewidth]{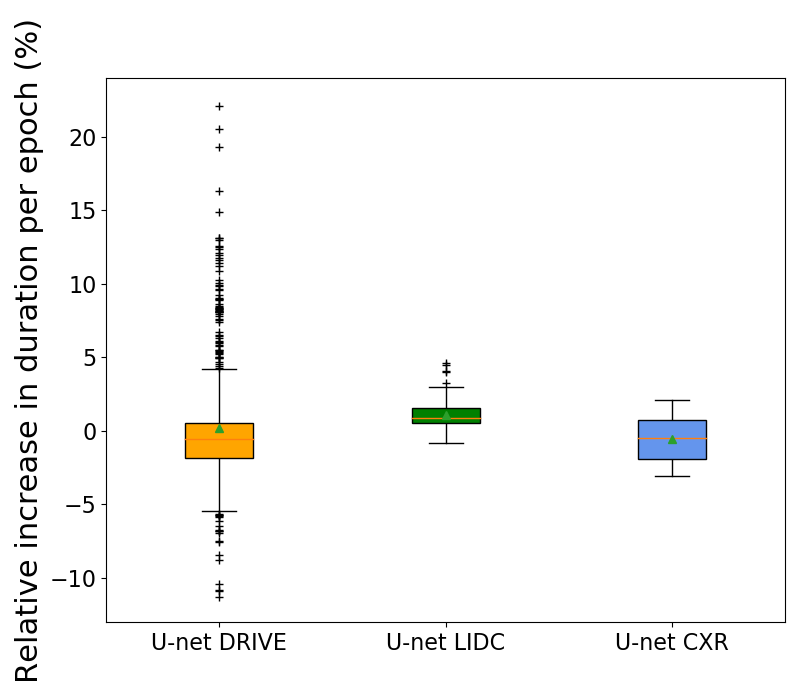}
    \caption{Box plot of the performance impact (\%) of using \textit{carbontracker} to monitor all training epochs shown as the relative increase in epoch duration compared to a baseline without \textit{carbontracker}. The whiskers and outliers are obtained from the Tukey method using 1.5 times IQR.}
    \label{fig:performance_impact_boxplot}
\end{figure}

\subsection{Performance Impact of Carbontracker}

The performance impact of using \textit{carbontracker} to monitor all training epochs is shown as a boxplot in \autoref{fig:performance_impact_boxplot}. We see that the mean increase in epoch duration for our U-net models across two runs is $0.19$\% on DRIVE, $1.06$\% on LIDC, and $-0.58$\% on CXR. While we see individual epochs with a relative increase of up to $5$\% on LIDC and even $22$\% on DRIVE, this is more likely attributed to the stochasticity in epoch duration than to \textit{carbontracker}. We further note that the fluctuations in epoch duration (\autoref{fig:performance_impact_boxplot}) are not caused by \textit{carbontracker}. These fluctuations are also witnessed in the baseline runs.

\begin{figure}[h!]
    \centering
    \includegraphics[width=0.9\linewidth]{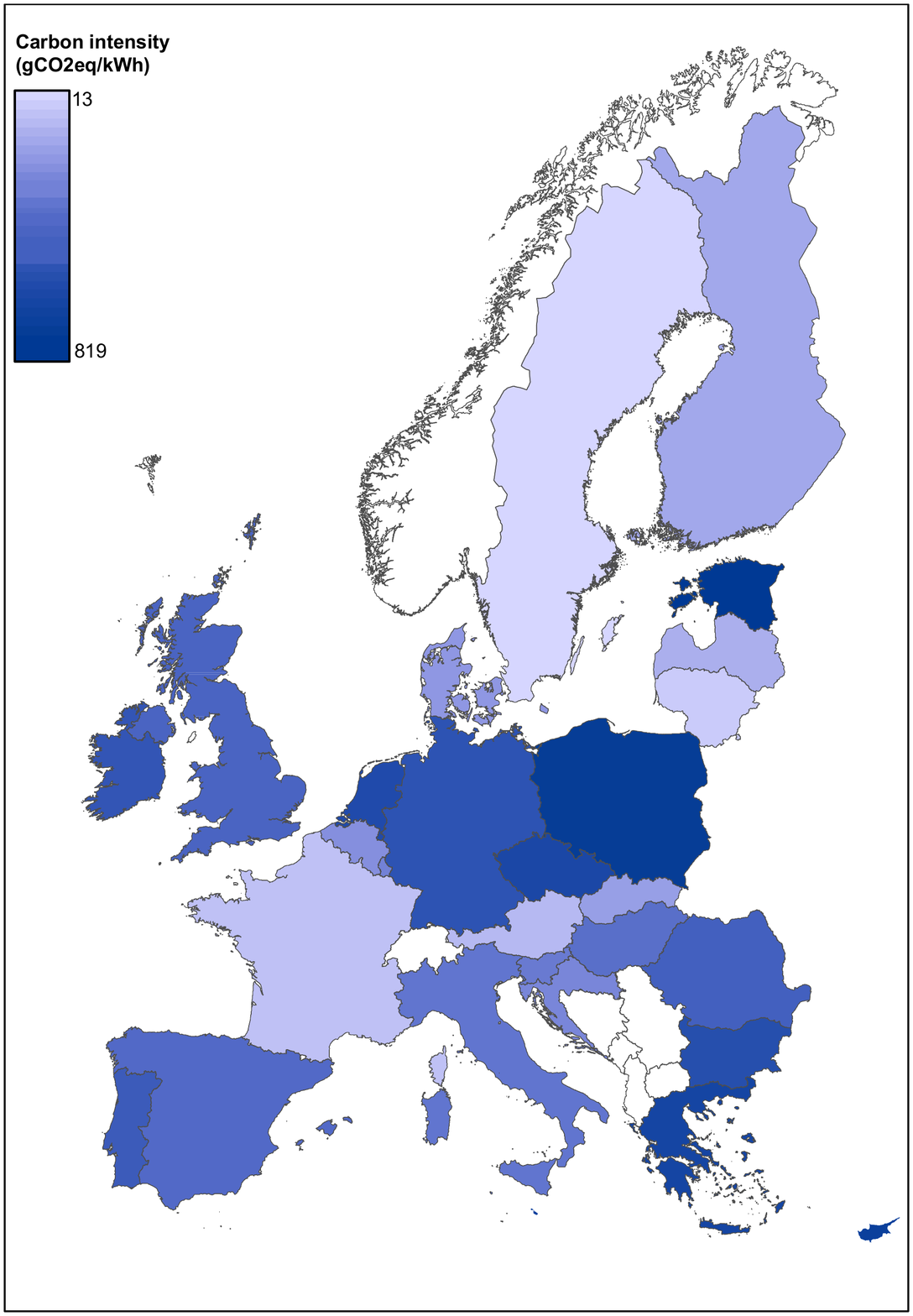}
    \caption{Average carbon intensity (\SI{}{\carbon}) of EU-28 countries in 2016. The intensity is calculated as the ratio of emissions from public electricity production and gross electricity production. Data is provided by the European Environment Agency (EEA). See \url{https://www.eea.europa.eu/ds_resolveuid/3f6dc9e9e92b45b9b829152c4e0e7ade}.}
    \label{fig:carbonintensity_europe}
\end{figure}

\begin{figure*}[h!]
    \centering
    \includegraphics[width=0.75\linewidth]{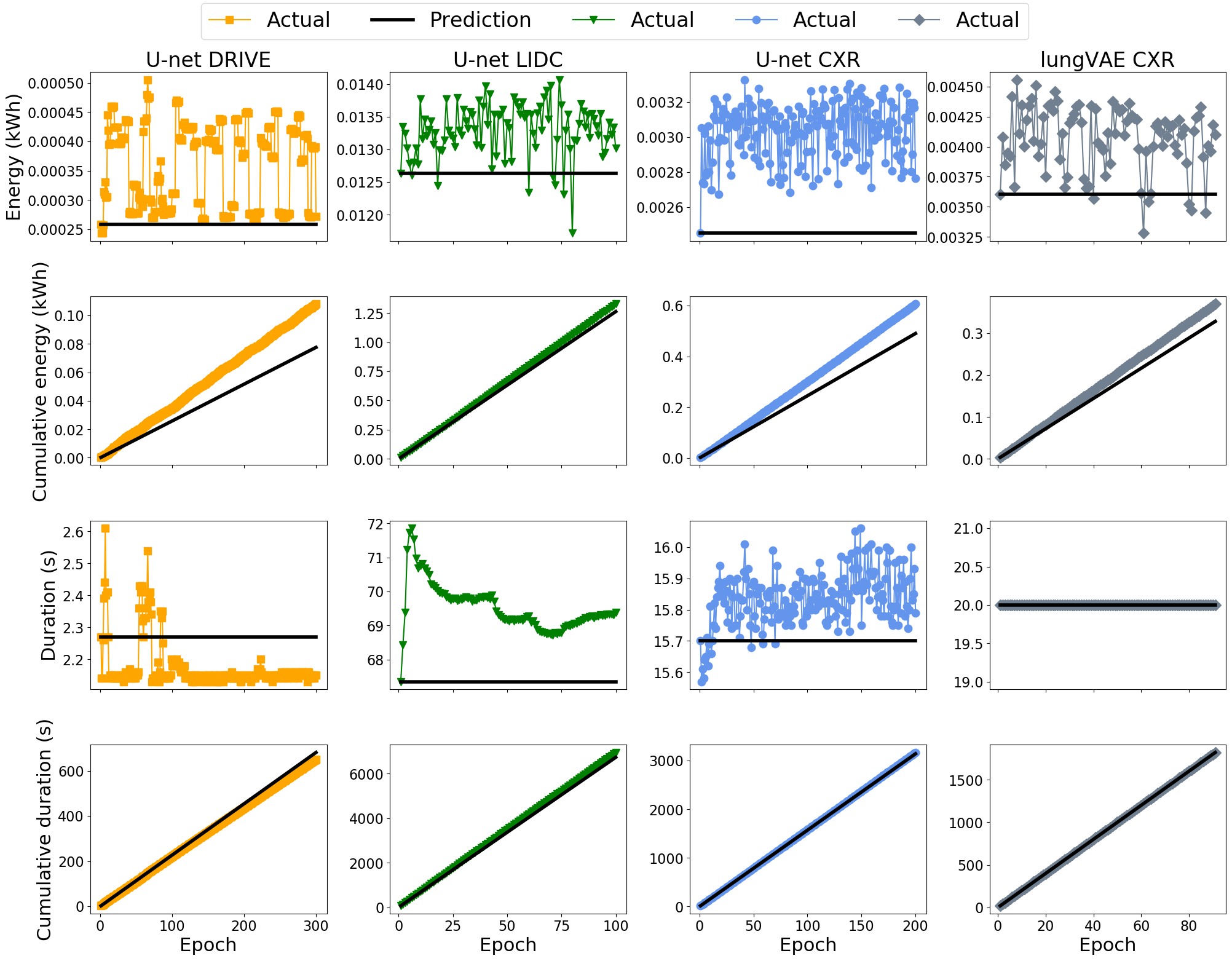}
    \caption{Comparison of predicted and measured values of energy (\SI{}{\kilo\watt\hour}) and duration (\SI{}{\second}) per epoch when predicting after a single epoch. (row 1) Energy. (row 2) Cumulative energy. (row 3) Duration. (row 4) Cumulative duration. Each column shows a different model and dataset, as detailed in \autoref{sec:models_and_data}. The initial epoch is often characterized by low energy usage and short epoch duration compared to the following epochs. Notwithstanding, the linear prediction model used by \textit{carbontracker} may still lead to reasonable predictions after monitoring a single epoch.}
    \label{fig:energy_time_cumulative}
\end{figure*}

\section{Estimating the Energy and Carbon Footprint of GPT-3}\label{sec:appendix_gpt3}

\citet{brown2020language} report that the GPT-3 model with 175 billion parameters used $3.14 \cdot 10^{23}$ \gls{fpos} of compute to train using NVIDIA V100 GPUs on a cluster provided by Microsoft. We assume that these are the most powerful V100 GPUs, the V100S PCIe model, with a tensor performance of $130$ TFLOPS\footnote{\url{https://bit.ly/2zFsOK2}} and that the Microsoft data center has a \gls{pue} of 1.125, the average for new Microsoft data centers in 2015\footnote{\url{http://download.microsoft.com/download/8/2/9/8297f7c7-ae81-4e99-b1db-d65a01f7a8ef/microsoft_cloud_infrastructure_datacenter_and_network_fact_sheet.pdf}}. The compute time on a single GPU is therefore 
\[\frac{3.14 \cdot 10
^{23} \text{ FPOs}}{130 \cdot 10^{12} \text{ FLOPS}} = \SI{2415384615.38}{\second} = \SI{27955.84}{\day}.\] This is equivalent to about $310$ GPUs running non-stop for $90$ days. If we use the \gls{tdp} of the V100s and the PUE, we can estimate that this used 
\begin{align*}
\SI{250}{\watt} \cdot \SI{2415384615.38}{\second} \cdot 1.125 &= \SI{679326923075.63}{\joule} \\
&= \SI{188701.92}{\kilo\watt\hour}.
\end{align*}
Using the average carbon intensity of USA in 2017 of \SI{449.06}{\carbon}\footnote{\url{https://www.eia.gov/tools/faqs/faq.php?id=74&t=11}}, we see this may emit up to
\begin{align*}
    \SI{449.06}{\carbon} &\cdot \SI{188701.92}{\kilo\watt\hour} \\
    &= \SI{84738484.20}{\gco} \\
    &= \SI{84738.48}{\kgco}.
\end{align*}
This is equivalent to 
\[\frac{\SI{84738484.20}{\gco}}{\SI{120.4}{\gco\per\kilo\meter}} = \SI{703808.01}{\kilo\meter} \] travelled by car using the average \si{\co} emissions of a newly registered car in the European Union in 2018\footnote{\url{https://www.eea.europa.eu/data-and-maps/indicators/average-co2-emissions-from-motor-vehicles/assessment-1}}.


\end{document}